\documentclass[12pt]{article}
\usepackage{graphicx}

\newcommand\pubnumber{}
\newcommand\pubdate{\today}

\def\sydney{School of Physics\\
The University of Sydney\\ NSW 2006, AUSTRALIA}
\def\support{\footnote{For the Belle Collaboration}}

\def\Title#1{\begin{center} {\Large #1 } \end{center}}
\def\Author#1{\begin{center}{ \sc #1} \end{center}}
\def\Address#1{\begin{center}{ \it #1} \end{center}}

\newcommand\pubblock{\rightline{\begin{tabular}{l} \pubnumber\\
         \pubdate  \end{tabular}}}
\newenvironment{Abstract}{\begin{quotation}  }{\end{quotation}}
\newenvironment{Presented}{\begin{quotation} \begin{center} 
             PRESENTED AT\end{center}\bigskip 
      \begin{center}\begin{large}}{\end{large}\end{center} \end{quotation}}

\def\beq{\begin{equation}}
\def\eeq#1{\label{#1}\end{equation}}
\def\eeqn{\end{equation}}

\def\beqa{\begin{eqnarray}}
\def\eeqa#1{\label{#1}\end{eqnarray}}
\def\eeqan{\end{eqnarray}}

\let\bar=\overbar

\def\Dslash{\not{\hbox{\kern-4pt $D$}}}
\def\dslash{\not{\hbox{\kern-2pt $\del$}}}

\def\msb{{\bar{\ssstyle M \kern -1pt S}}}

\begin{document}
\begin{titlepage}
\pubblock

\vfill
\Title{Measurement of $B^0 \to \pi^{-} \ell^{+} \nu_\ell$ in untagged events and
determination of $|V_{ub}|$ at Belle}
\vfill
\Author{ Kevin Varvell\support}
\Address{\sydney}
\vfill
\begin{Abstract}
A study of the charmless semileptonic decay $B^{0} \to \pi^{-} \ell^{+} \nu_{\ell}$ 
using a large sample of untagged $\Upsilon(4S) \to B \bar{B}$ events collected with the Belle detector at the KEKB asymmetric $e^{+} e^{-}$ collider is presented. The
branching fraction of the decay is obtained and from this the Cabibbo-Kobayashi-Maskawa matrix element 
$|V_{ub}|$ is determined using two different approaches.
\end{Abstract}
\vfill
\begin{Presented}
6th International Workshop on the \\CKM Unitarity Triangle, CKM2010\\
University of Warwick, UK\\  September 6--10, 2010
\end{Presented}
\vfill
\end{titlepage}
\def\thefootnote{\fnsymbol{footnote}}
\setcounter{footnote}{0}

\section{Introduction}

The branching fraction for the exclusive weak decay $B^0 \to \pi^- \ell^+ \nu$ provides an attractive means for extracting the Cabibbo-Kobayashi-Maskawa (CKM) matrix element $\arrowvert V_{ub} \arrowvert$. The extraction method requires knowledge of the form factor $f_+(q^2)$ for the decay, where $q^2$ is the square of the four-momentum transfer to the charged lepton-neutrino system. Traditionally this input comes from theoretical calculations such as those based on lattice QCD \cite{Dalgic:2006dt} \cite{Okamoto:2004xg}, light cone sum rules \cite{Ball:2004ye} or a relativistic quark model \cite{Scora:1995ty}. The present study obtains the differential branching fraction for this decay in 13 regions of $q^2$ and from this extracts $\arrowvert V_{ub} \arrowvert$ using the above theoretical predictions. It also employs a variation on this approach which has appeared more recently \cite{Boyd:1994tt} \cite{Bailey:2008wp}, based on a simultaneous fit to data and lattice predictions, from which a value for $\arrowvert V_{ub} \arrowvert$ emerges in a more model-independent way.
A more extensive description of the study presented here is available through the arXiv e-print server~\cite{Ha:2010rf}.

\section{Extraction of the branching fraction}

The data sample consists of $657 \times 10^6$ $B \overline{B}$ pairs collected by the Belle experiment~\cite{:2000cg} at the KEKB asymmetric $e^+ e^-$ collider operating at the $\Upsilon(4S)$ resonance. $B^0 \to \pi^- \ell^+ \nu$ decays are reconstructed from pairs of oppositely charged leptons (muon or electron) and pions, with the neutrino reconstructed from the missing 3-momentum in the center of mass frame (CM); $p_\nu = \left( \arrowvert \vec{p}_{\mathrm{miss}} \arrowvert, \vec{p}_{\mathrm{miss}} \right)$, where $\vec{p}_{\mathrm{miss}} = - \Sigma_i \vec{p}_i$ and the sum extends over all observed particles in the event. Signal event selection uses the beam-energy-constrained mass $M_{\mathrm{bc}} = \sqrt{E_{\mathrm{beam}}^2 - {\arrowvert \vec{p}_\pi + \vec{p}_\ell + \vec{p}_\nu \arrowvert}^2}$ and $\Delta E = E_{\mathrm{beam}} - \left( E_\pi + E_\ell + E_\nu \right)$, where $E_{\mathrm{beam}}$ is the beam energy in the CM. The conditions $M_{\mathrm{bc}} > 5.19$\ GeV/$c^2$ and $\arrowvert \Delta E \arrowvert < 0.1$\ GeV are imposed. Backgrounds are estimated using Monte Carlo samples of other $B \to X_u \ell \nu$ decays, $B \to X_c \ell \nu$ decays and continuum.

\begin{figure}[htp]
\begin{center}
\begin{tabular}{cccc}
\includegraphics[scale=0.25]{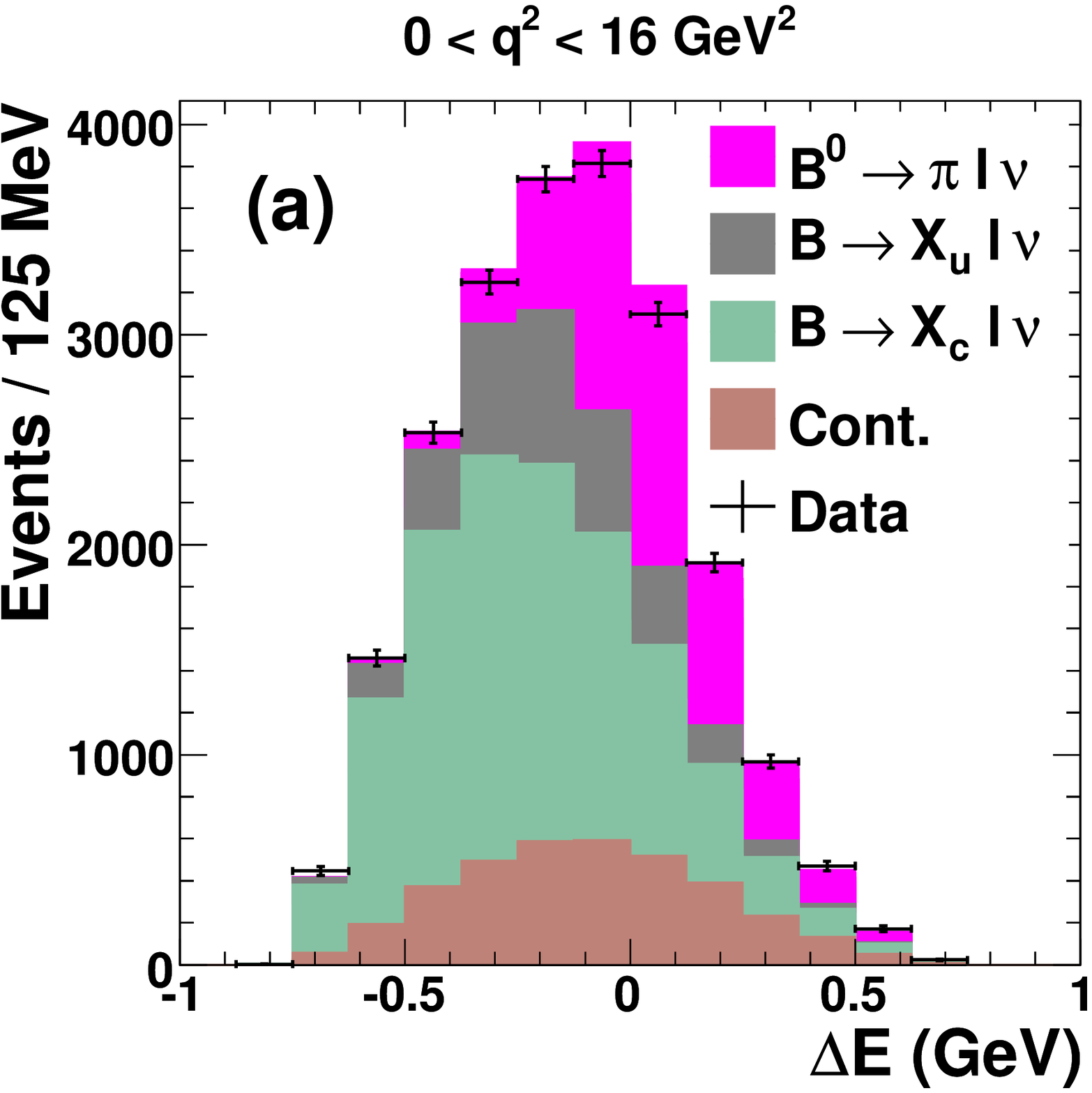}   &
\includegraphics[scale=0.25]{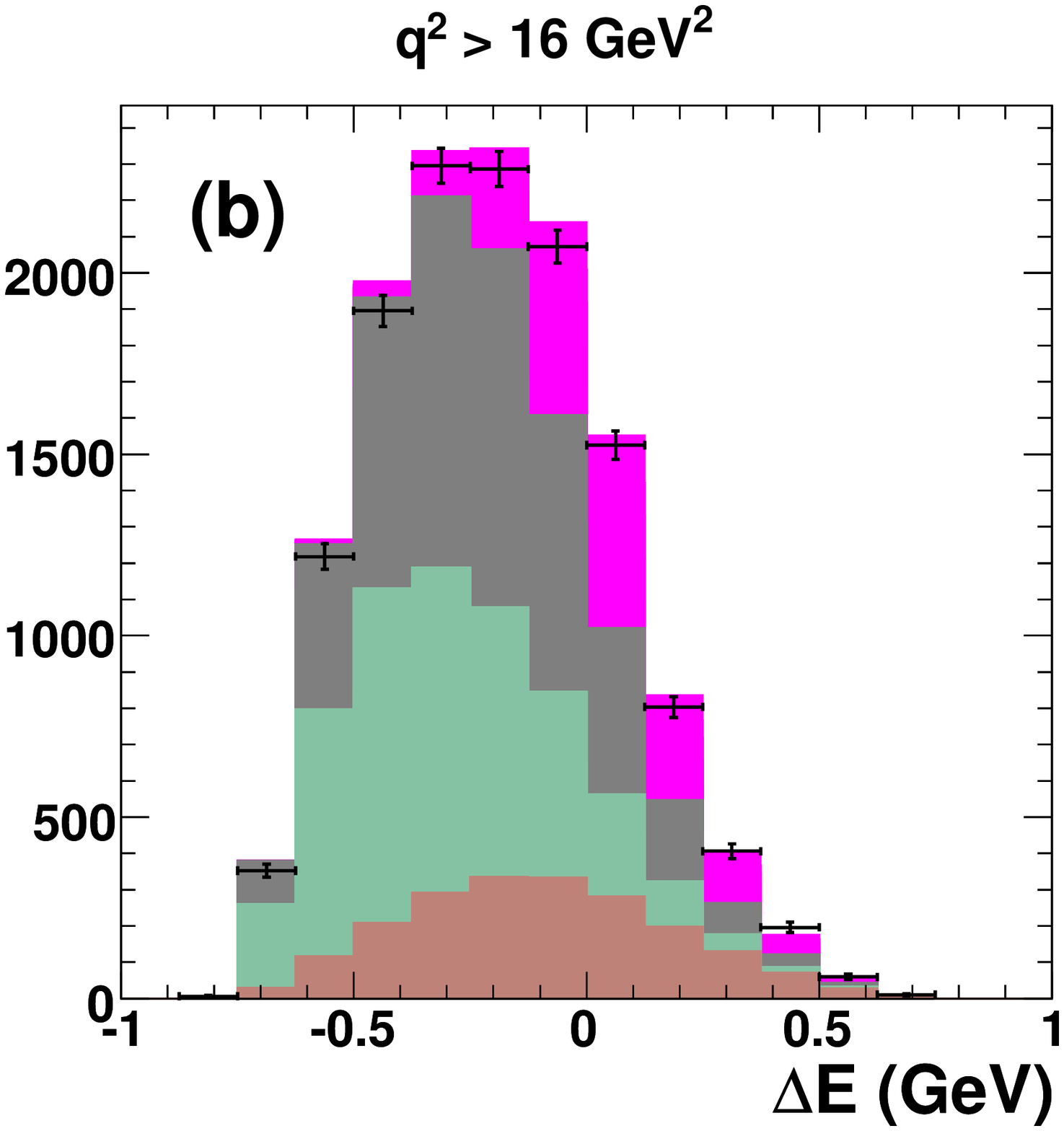}  \\
\includegraphics[scale=0.25]{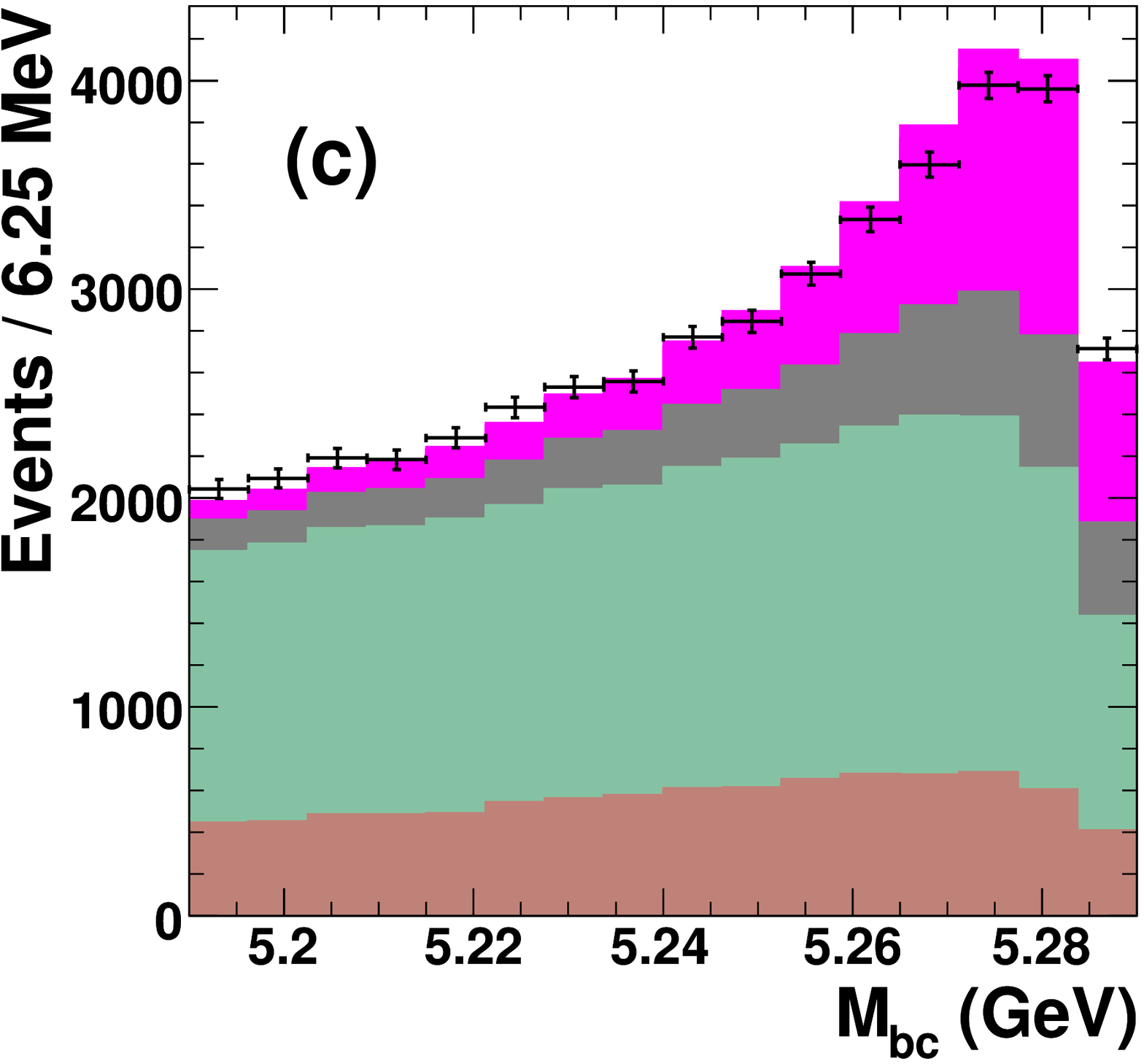}  &
\includegraphics[scale=0.25]{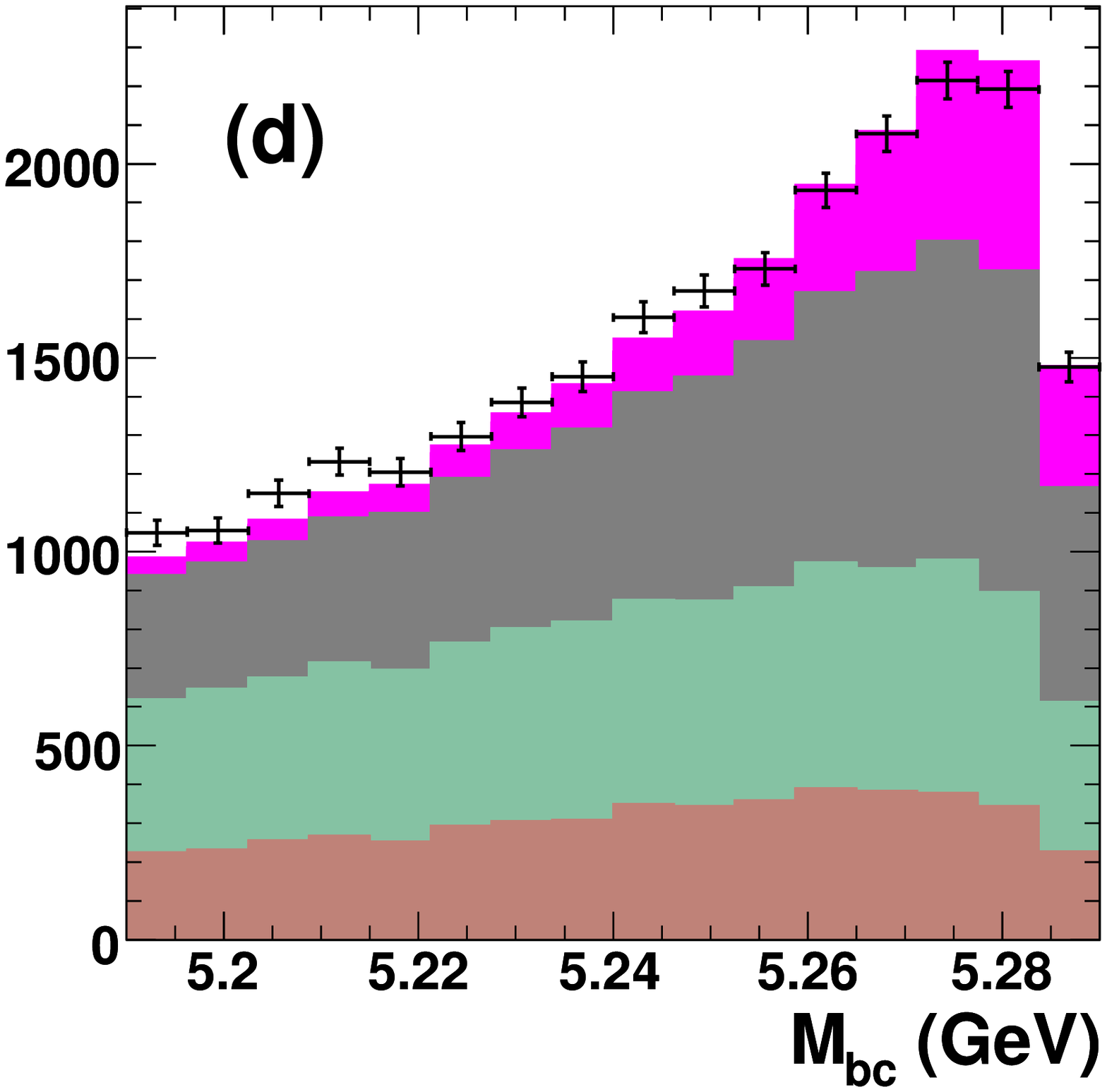} \\
\end{tabular}
\end{center}
\caption{Fit projections in $\Delta E$ with $M_{\mathrm{bc}} > 5.27$ GeV/$c^2$ for low $q^2 < 16$\ GeV$^2/c^2$ (a) and high $q^2 > 16$\ GeV$^2/c^2$ (b) regions, and corresponding projections in $M_{\mathrm{bc}}$ with $\arrowvert \Delta E \arrowvert < 0.125$\ GeV in (c) and (d). Data points show statistical errors only. The histograms show, from top to bottom, signal $B^0 \to \pi^- \ell^+ \nu$, $B \to X_u \ell \nu$, $B \to X_c \ell \nu$ and continuum components.}
\label{fig:de_mbc}
\end{figure}

\begin{figure}[hbp]
\begin{center}
\includegraphics[scale=0.35]{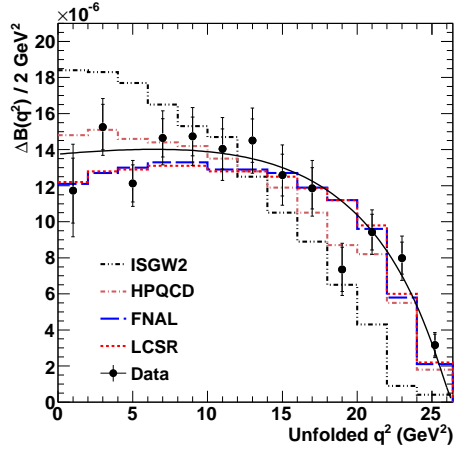}
\end{center}
\caption{The partial branching fraction as a function of $q^2$. Data points (closed circles) indicate the statistical uncertainties (inner bars) and the combined statistical and systematic uncertainties (outer bars). The histograms show predictions based on various form factor prescriptions. The curve is the result of a fit to the data using the BK form factor parameterization \cite{Becirevic:1999kt}.}
\label{fig:bf}
\end{figure}

A binned extended likelihood fit is used to extract the signal, in the two-dimensional $\left( M_{\mathrm{bc}}, \Delta E \right)$ plane and in 13 bins of $q^2$. Projections of the fit are shown in Fig.~\ref{fig:de_mbc}. After the $q^2$ distribution of the signal events is unsmeared and corrected for signal efficiency, the distribution of the partial branching fraction is
calculated, with the result displayed in Fig.~\ref{fig:bf}. Of the four models which are compared to the partial branching fraction distribution in Fig.~\ref{fig:bf}, the ISGW2 model is excluded by the data.

Summed over the complete range in $q^2$, the branching fraction for the decay is determined to be $\mathcal{B} \left( B^{0} \to \pi^{-} \ell^{+} \nu_{\ell} \right) = \left( 1.49 \pm 0.04 \left( {\rm stat.} \right) \pm 0.07 \left( {\rm syst.} \right) \right) \times 10^{-4}$.


\begin{table}[htp]
\begin{center}
\begin{tabular}{lccc}
\hline
\hline
                             & $q^2$ (GeV$^2/c^2$) & $\Delta \zeta$ (ps$^{-1}$) & $\vert V_{ub} \vert$(10$^{-3}$)          \\
\hline
HPQCD~\cite{Dalgic:2006dt}   &  $>$ 16         & 2.07 $\pm$ 0.57            & 3.55 $\pm$ 0.09 $\pm$ 0.09 $^{+0.62}_{-0.41}$ \\
FNAL~\cite{Okamoto:2004xg}   &  $>$ 16         & 1.83 $\pm$ 0.50            & 3.78 $\pm$ 0.10 $\pm$ 0.10 $^{+0.65}_{-0.43}$ \\
LCSR~\cite{Ball:2004ye}      &  $<$ 16         & 5.44 $\pm$ 1.43            & 3.64 $\pm$ 0.06 $\pm$ 0.09 $^{+0.60}_{-0.40}$ \\
ISGW2~\cite{Scora:1995ty}    &  all            &  9.6 $\pm$ 4.8             & 3.19 $\pm$ 0.04 $\pm$ 0.07 $^{+1.32}_{-0.59}$ \\
\hline
\hline
\end{tabular}
\caption{Values for $|V_{ub}|$ extracted using different form factor predictions. The quoted uncertainties in $|V_{ub}|$ 
are statistical, systematic and arising from the uncertainty in $\Delta \zeta$, respectively.}
\label{tab:vub}
\end{center}
\end{table}
\begin{figure}[htp]
\begin{center}
\includegraphics[scale=0.40]{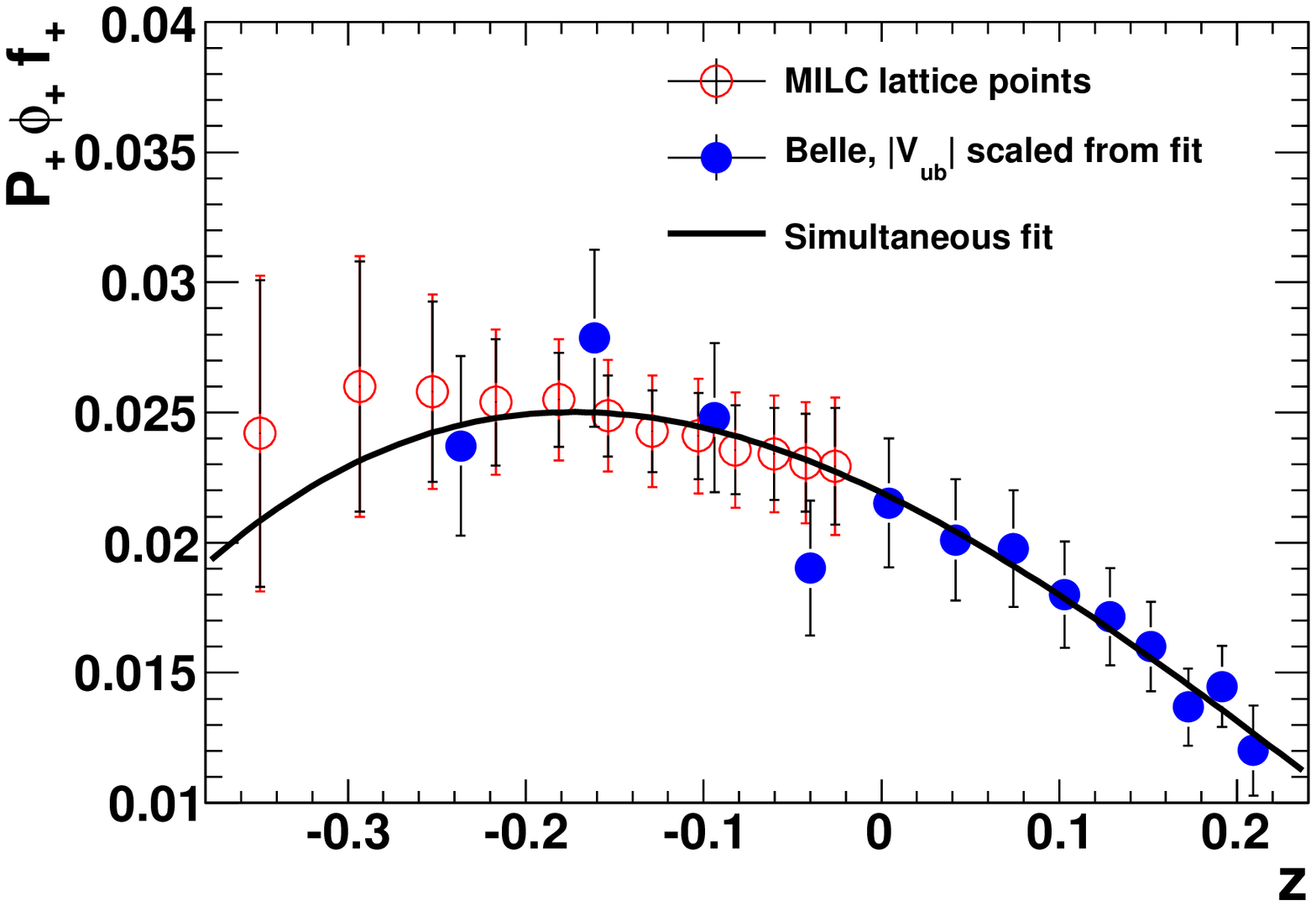}
\end{center}
\caption{The result of a simultaneous fit to the present experimental data points (closed circles, blue) and FNAL/MILC lattice QCD points~\cite{Bailey:2008wp} (open circles, red) used to extract $|V_{ub}|$. The data points have been scaled by the value of $|V_{ub}|$ obtained from the fit. For the data points both statistical and systematic uncertainties are combined; for the lattice points the inner bars show statistical uncertainties and the outer bars combined statistical and systematic uncertainties. The functions $P_+$ and $\phi_+$ are described in reference~\cite{Bailey:2008wp}.}
\label{fig:fit}
\end{figure}

\section{Determination of $|V_{ub}|$}

$\arrowvert V_{ub} \arrowvert$ can be extracted from the partial branching fraction information through the relation $\vert V_{ub} \vert = \sqrt{ \Delta \mathcal{B}(q^2) / ( \tau_{B^0} \Delta \zeta ) }$, where $\tau_{B^0} = 1.525 \pm 0.009$\ ps is the $B^0$ lifetime. The normalised partial decay rates $\Delta \zeta$ are predicted by theory. The results, which are dominated by theoretical uncertainties, are summarised in Table~\ref{tab:vub} for several sources of input for $\Delta \zeta$.

A model-independent determination of $\arrowvert V_{ub} \arrowvert$ is obtained by simultaneously fitting the branching fraction data and the MILC lattice QCD form-factor after transforming to the so-called ``z-parameterization''~\cite{Bailey:2008wp}. The result of the fit is shown in Fig.~\ref{fig:fit}, with the result $\arrowvert V_{ub} \arrowvert = \left( 3.43 \pm 0.33 \right) \times 10^{-3}$, where the uncertainty combines both statistical and systematic contributions.

\section{Conclusions}

The branching fraction $\mathcal{B} \left( B^{0} \to \pi^{-} \ell^{+} \nu_{\ell} \right) = \left( 1.49 \pm 0.04 \left( {\rm stat.} \right) \pm 0.07 \left( {\rm syst.} \right) \right) \times 10^{-4}$ has been measured and from this $\arrowvert V_{ub} \arrowvert$ has been determined, using both model-dependent and model-independent approaches. The latter approach provides the best overall precision, producing a value $\arrowvert V_{ub} \arrowvert = \left( 3.43 \pm 0.33 \right) \times 10^{-3}$ where the uncertainty incorporates both statistical and systematic effects.


\end{document}